\pdfoutput=1

\documentclass{elsarticle}
\usepackage{graphicx}
\usepackage{dcolumn}
\usepackage{bm}
\usepackage{gensymb}

\usepackage{amsmath, amssymb,amsfonts}
\newenvironment{affiliations}{%
    \setcounter{enumi}{1}%
    \setlength{\parindent}{0in}%
    \slshape\sloppy%
    \begin{list}{\upshape$^{\arabic{enumi}}$}{%
        \usecounter{enumi}%
        \setlength{\leftmargin}{0in}%
        \setlength{\topsep}{0in}%
        \setlength{\labelsep}{0in}%
        \setlength{\labelwidth}{0in}%
        \setlength{\listparindent}{0in}%
        \setlength{\itemsep}{0ex}%
        \setlength{\parsep}{0in}%
        }
    }{\end{list}\par\vspace{12pt}}

\newenvironment{methods}{%
    \section*{Methods}%
    \setlength{\parskip}{12pt}%
    }{}

\newenvironment{addendum}{%
    \setlength{\parindent}{0in}%
    \small%
    \begin{list}{Acknowledgements}{%
        \setlength{\leftmargin}{0in}%
        \setlength{\listparindent}{0in}%
        \setlength{\labelsep}{0em}%
        \setlength{\labelwidth}{0in}%
        \setlength{\itemsep}{12pt}%
        }
    }
    {\end{list}\normalsize}

\title{Monoenergetic proton beams accelerated by a radiation pressure driven shock}

\author{
Charlotte A.~J.~Palmer$^{1}$, 
N.~P.~Dover$^{1}$, 
I.~Pogorelsky$^{2}$,
M.~ Babzien$^{2}$,
G.~I.~Dudnikova$^{3}$
M.~Ispiriyan$^{4}$,
M.~N.~Polyanskiy$^{2}$,
J.~Schreiber$^{1,5,6}$,
P.~Shkolnikov$^{4}$,
V.~Yakimenko$^{2}$,
Z.~Najmudin$^{1}$
}

\begin{document}

\begin{abstract}
High energy ion beams ($>$ MeV) generated by intense laser pulses promise to be viable alternatives to conventional ion beam sources due to their unique properties such as high charge \cite{Clark:2000kx,Snavely:2000bv}, low emittance \cite{Borghesi:2004fk,Cowan:2004uq}, compactness and ease of beam delivery \cite{Bulanov:2002zr}.  Typically the acceleration is due to the rapid expansion of a laser heated solid foil, but this usually leads to ion beams with large energy spread. Until now, control of the energy spread has only been achieved at the expense of reduced charge and increased complexity \cite{Schwoerer:2006ec, Hegelich:2006dq, Toncian:2006fu}. Radiation pressure acceleration (RPA) provides an alternative route to producing laser-driven monoenergetic ion beams \cite{Esirkepov:2004kl, Robinson:2008mi}. In this paper, we show the interaction of an intense {\em infrared} laser with a {\em gaseous} hydrogen target can produce proton spectra of small energy spread ($\sigma \sim 4\%$), and low background. The scaling of proton energy with the ratio of intensity over density ($I/n$) indicates that the acceleration is due to the shock generated by radiation-pressure driven hole-boring of the critical surface \cite{Wilks:1992ys, Zepf:1996ve}. These are the first high contrast mononenergetic beams that have been theorised from RPA \cite{Esirkepov:2004kl,Robinson:2008mi, Liseykina:2008cr,Macchi:2009qa,Qiao:2009oz}, and makes them highly desirable for numerous ion beam applications.
 
\end{abstract}

\maketitle

\begin{affiliations}
\item{ The Blackett Laboratory, Imperial College London, SW7 2BW, United Kingdom}
\item{ Accelerator Test Facility, Brookhaven National Laboratory, NY 11973, USA}
\item{University of Maryland, College Park, MD 20742, USA}
\item{Stony Brook University, Stony Brook, NY 11794, USA}
\item{ Fakult\"at f\"ur Physik, Ludwig-Maximilians-Universit\"at M\"unchen, D-85748 Garching, Germany}
\item{ Max-Planck-Institut f\"ur Quantenoptik, Hans-Kopfermann-Str. 1, D-85748 Garching, Germany}

\end{affiliations}

Most investigations of ion acceleration by intense lasers have relied on sheath acceleration. When a high-intensity laser impacts on a solid target, an overdense (opaque) plasma is formed, i.e.~$n_{e} > n_{crit} = \epsilon_0m\omega_{0}^{2}/e^{2}$. Laser energy is converted into hot electrons with temperature $k_{B}T_{e}$, which expand to form an electrostatic sheath around the target with fields of order $\sim {\rm MV}/\mu$m. This sheath accelerates surface ions to energies $\approx k_{B}T_{e}$ ($\sim$ multi-MeV) per nucleon  \cite{Mora:2003dp,Wilks:2001ai}.
However, this usually produces ion beams of broad energy spread, with a large number of ion species, many originating from impurities \cite{Clark:2000qe,Snavely:2000bv}. Also, the production of proton beams has until now mostly depended on the presence of the same target impurities (hydrocarbons and moisture). Due to the bulk heating, ions are accelerated from both front and back surfaces, providing inefficient energy coupling between the laser and the ions, and ion energy scales only weakly with irradiance ($\sim (I\lambda^{2})^{1/2}$) \cite{Fuchs:2006fk,Schreiber:2006uq}. 
Modulated ion energy spectra have been produced, either by using complex target preparation \cite{Hegelich:2006dq,Schwoerer:2006ec}, or energy selection \cite{Toncian:2006fu}, but these methods still feature relatively poor peak-to-noise spectral contrast, reduced yields and the presence of impurity species. 

Radiation pressure acceleration (RPA) has been proposed as an alternative method of ion acceleration at ultrahigh intensities \cite{Esirkepov:2004kl,Robinson:2008mi, Liseykina:2008cr,Macchi:2009qa,Qiao:2009oz}. For an opaque plasma ($n > n_{cr}$), the laser radiation pressure, $P_{R} = 2I/c$, initially pushes plasma electrons into the target setting up an electrostatic shock whose space charge field pulls along ions at the hole-boring velocity \cite{Wilks:1992ys,Zepf:1996ve}, $v_{hb} \approx (2I/\rho c)^{1/2}$, where $I$ is the intensity and $\rho$ is the mass density. Stationary ions in advance of the shock front can be accelerated by the same space charge field, effectively bouncing off the shock front, producing a population of energetic ions with velocity $2 v_{hb}$ \cite{Denavit:1992uq, Silva:2004lh, Wei:2004fk}. If the target becomes sufficiently thin that all of the electrons can be pushed out of the target ($d < c/\omega_{p}$), the ions can be pulled along in unison, in what is called the ``light-sail'' phase of RPA. The ``light-sail'' phase is necessary to reach higher energy \cite{Qiao:2009oz, Macchi:2009qa}. First experiments have now reported limited enhancement of ion energy contrast with optical lasers and ultrathin solid targets \cite{Henig:2009fk}. However, due to the extremely high intensities required, energy gains were modest, energy spreads large, and impurity-free proton beams could not be produced. 

Gas targets have been shown to be an alternative to solid foils. They  operate at high repetition rate, are easily adjusted for target density and material and are less prone to contamination \cite{Willingale:2006kb}. However, they are difficult to operate at near or above critical density, which is necessary for efficient ion acceleration (e.g.~for $1 \, \mu$m laser, $n_{crit} \simeq 1\times 10^{21}$ cm$^{-3}$) \cite{Esirkepov:2006ys, Willingale:2008vn}.
This can be remedied by use of longer wavelength (infrared) lasers. For example, for a $\lambda \approx 10 \, \mu$m CO$_{2}$ laser, $n_{crit} \ (\propto \lambda^{-2}) \approx 10^{19}$ cm$^{-3}$.  This density is easily obtained by ionisation of gas targets.  
Importantly, due to the lower specific mass ($<10^{4} \times$ solid densities), gas targets become of interest for RPA at proportionally reduced laser intensity.

\begin{figure}[tb]
\begin{center}
\includegraphics[scale=0.45]{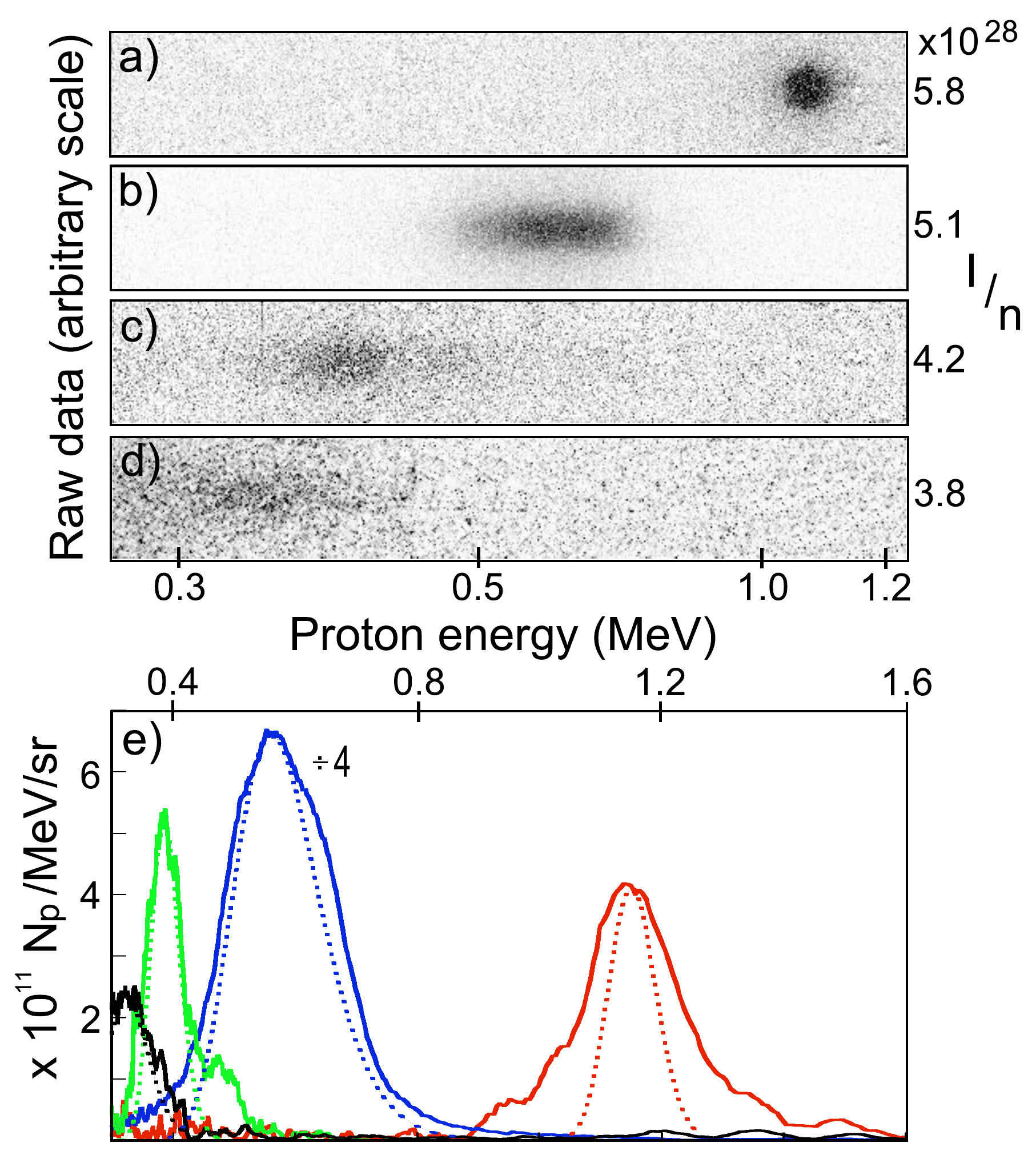}
\caption{{\bf Ion spectra:} Selection of raw and processed proton spectra for varying peak density ($n$) and vacuum intensity ($I$) showing scaling of peak proton energy $E_{max} \propto I /n$.  Parameter $I /n$ shown to the right of the respective raw images. Shots taken with; 
a) $I = 7.5$,~$n=6.1 n_{cr}$, b) $I = 6.5$,~$n =6.1 n_{cr}$, c) $I = 6.9$,~$n =7.6n_{cr}$, d) $I = 6.7$,~$n =8.0n_{cr}$  (intensities in units $10 ^{15}\,  {\rm Wcm}^{-2}$). e) Background subtracted (solid lines) and also corrected (dashed lines) spectra.  Heights of corrected spectra adjusted to match those of raw line-outs. Line-out corresponding to b) reduced $4\times$ to fit on same scale.
}\label{spectra}
\end{center}
\end{figure}

An experiment designed to investigate radiation driven acceleration of gas targets by an infrared ($\lambda = 10\, \mu$m) laser at intensities up to $10^{16}\, {\rm Wcm}^{-2}$ was performed (see methods). The interaction was diagnosed by transverse probing and measurements of the forward accelerated ion beam. A gas nozzle of diameter $L = 1$~mm was employed, which though only $L\approx 100 \lambda $, is not thin enough for ``light-sail'' RPA.  The normalised vector potential was $a_{0}\approx 0.5$  (see methods).  Since the ponderomotive potential $U_{p} = a_{0}^{2}m_{e}c^{2}$, for this $a_{0}$ one would expect to produce hot electrons, which would promote sheath acceleration. However, this was mitigated to some extent by use of circular polarisation in the experiment \cite{Macchi:2005kl, Robinson:2008mi}. 

Figure 1 shows a selection of ion beams recorded with the magnetic spectrometer for a range of densities and intensities. Ion beams with a peak energy up to $E_{max}  =1.1$ MeV were observed. The most striking observation is the narrow spectral width of the ion feature, especially at higher $I/n$. The peak-to-noise contrast of the spectrum is greater than 100. 
In figure 1a, the ion beam image is a  circle comparable in size to the projected aperture size at the image plane, indicating that this feature is dominated by the instrument function of the spectrometer. Spectra (figure 1e), taken after deconvolving the instrument response (see methods), show that the narrowest observed feature (figure 1a) corresponds to an {\em rms} energy spread of only $\sigma = 4.2~\%$. 

For the shot peaking at $0.6$ MeV (figure 1b), the maximum of the spectrum is $\sim 3 \times 10^{12}$ protons/MeV/sr, which is up to 1000 times greater than for previous modulated ion beams from laser plasma interactions \cite{Schwoerer:2006ec, Hegelich:2006dq,Henig:2009fk}. From the transverse spread of the $4\times10^{6}$ protons passing through the spectrometer pinhole, the geometrical emittance of the beam was determined to be $\epsilon = 0.16$ mm-mrad, corresponding to a normalised emittance of $\epsilon_{n} \equiv \beta \gamma \epsilon = 8$ nm-rad.

Also apparent is that the peak energy increases with increasing $I/n$ over this density range. This is shown more clearly in figure 2, which plots measured peak energy versus expected energy due to shock acceleration, $E = \frac{1}{2} m (2v_{hb})^{2} = 4I/n c$. The data shows a trend of increasing energy for increasing $I/n$.  However, the linear best fit implies an  $I /n$ which is 9 times higher than expected by taking the vacuum focused intensity, and the peak measured densities. This discrepancy will be addressed further below.

\begin{figure}
\begin{center}
\includegraphics[scale=0.40]{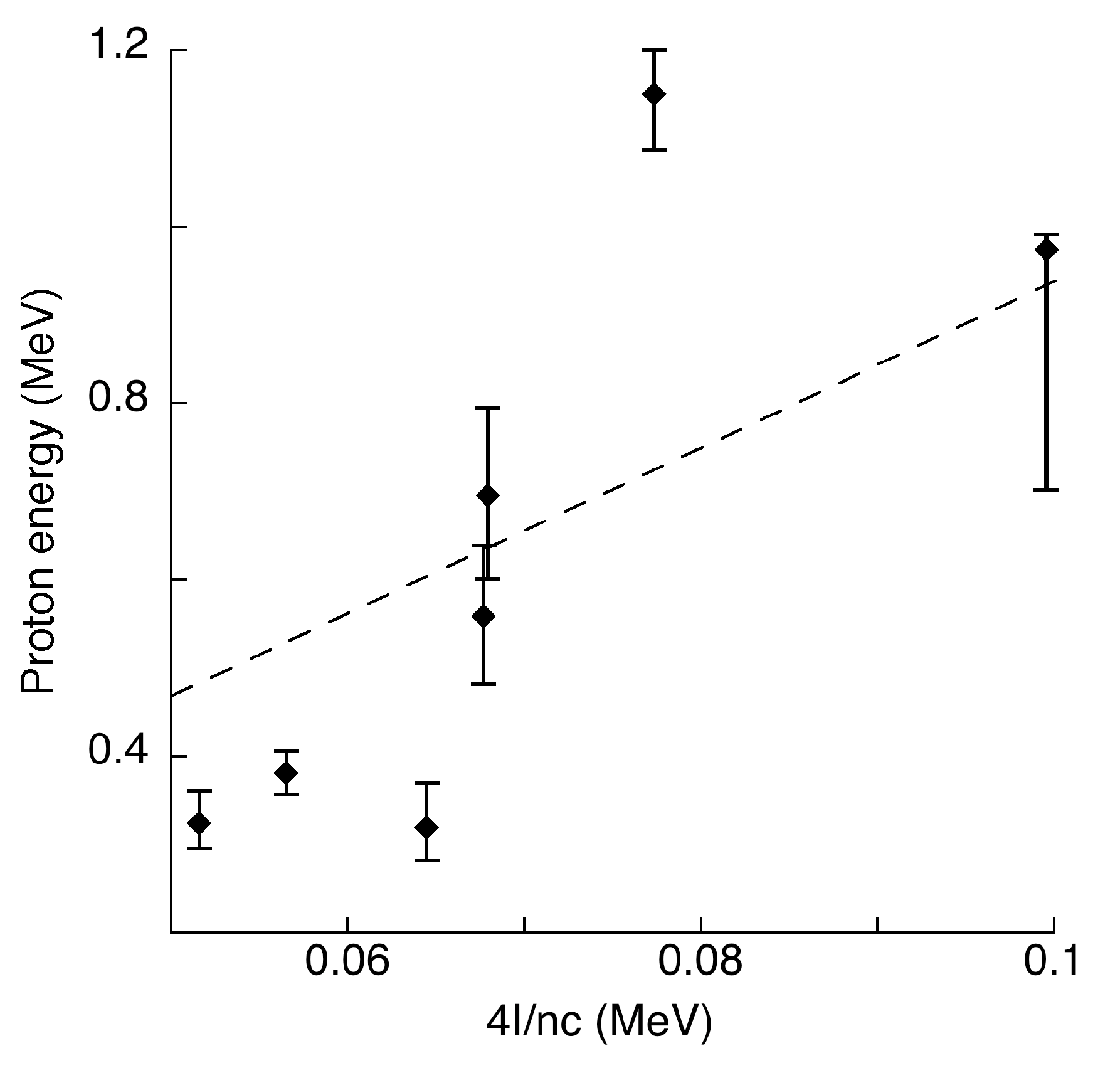}
\caption{{\bf Proton energy scaling:} as a function of expected energy due to hole-boring, $E = 4I/nc$. For $8 n_{cr} > n > 4 n_{cr}$, peak proton signal increases with increasing $I/n$. For $n < 4 n_{cr}$ beams were produced with larger variation in energy (below the observed scaling), and are not included. No beams observed for $n > 8 n_{cr}$. Vertical bars represent rms energy spread of each shot.}
\label{default}
\end{center}
\end{figure}

Simultaneous transverse optical probing was also performed. Although opaque to the infrared driver, at these densities the plasma is transparent at optical wavelengths.  By contrast, probing the surface of a solid-density interaction would be impossible due to severe refraction. Figure 3a shows a shadowgram of an interaction for peak density $n = 4 n_{cr}$, at $t \sim 30$~ps after arrival of the main CO$_{2}$ pulse, shortly after the end of the interaction with the second pulse.  A large cavity has been created by the laser inside the gas target. Interferometry (figure 3b) shows that the plasma within the cavity is at much lower density ($< 10^{18} \, {\rm cm}^{-3}$), whilst the walls of the cavity are just above critical density. Hence this density-discontinuity (shock) is associated with hole-boring. The laser has travelled up to the critical surface, at which point it is mostly reflected. The resultant radiation pressure causes the critical surface to be driven inwards.  Probing at later times shows the shock front moving further into the target, though at reduced velocity due to energy dissipation. Features are seen inside the overdense region of the interaction due to the hot electrons created at the shock front.  At later times, it is possible to see a slower evolution of the back surface of the gas target, which was presumably not detected by the ion diagnostics since the associated energy is much lower. 

To model the interaction, a series of 2D particle-in-cell simulations were performed. A hole-boring front can be seen (figure 4a) forming above critical density, but well below the peak density.  The velocity of the front increases up to the peak of the laser pulse where it is comparable to the expected hole-boring velocity, $v_{hb}$.   As the peak density is varied in the simulations, the velocity of the hole-boring front is found to decrease. This is because the shock, though starting near critical density, reaches higher density as it moves forward.  Secondary pulses enhance the shock structure and snowplough a greater number of protons to $v_{hb}$, allowing protons below the hole-boring velocity to ``catch-up'' with the shock front. This can serve to improve the fidelity of the shock structure.  

The simulations also show self-focusing of the laser pulses (by a factor $>2$ in intensity)\cite{Sun:1987kx}. This in conjunction with the lower than peak interaction density explains the previously noted discrepancy between measured and predicted ion energies in the experiment. 
Self-focusing would be accentuated in three dimensions. To account for this a further simulation was performed with $a_0=0.9$ and shorter density scale-length.  Figure 4b shows the spectrum observed in a simulated spectrometer, which models the experimental one.  A clear peak is obseved at high energy with similar energy to that observed experimentally.  Assuming a real beam divergence comparable to that of the simulated high energy proton bunch ($\sim 4^{\circ}$) would imply a total number of accelerated protons in the bunch of up to $5\times10^{9} ~(\simeq 0.8$ nC). Such a high-current low energy spread beam of protons would obviously be of interest for applications as diverse as hadrontherapy \cite{Bulanov:2002zr} and as an injector for high-energy particle accelerators.

\begin{figure}
\begin{center}
\includegraphics[scale=0.3]{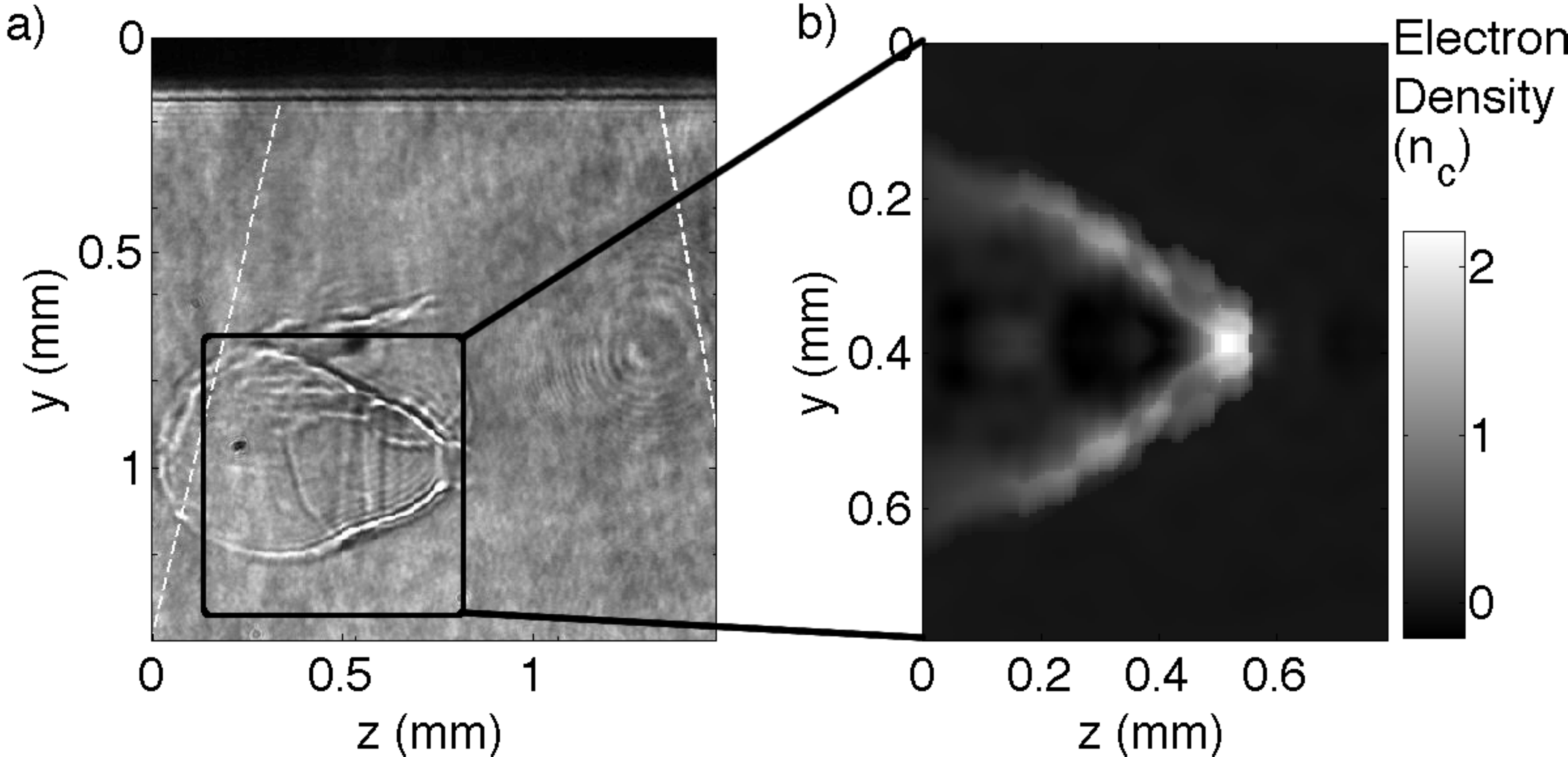}
\caption{{\bf Optical probing:} (a) Shadowgraphy of the interaction for $n = 4 n_{cr}$, 30ps after the first main pulse interacts with the gas.  Laser entered from left, with the silhouette of the gas nozzle shown above. White dotted lines approximate initial boundaries of the gas jet.  (b) Density profiles obtained from simultaneous interferogram}
\label{default}
\end{center}
\end{figure}

\begin{figure}
\begin{center}
\includegraphics[scale=1.2]{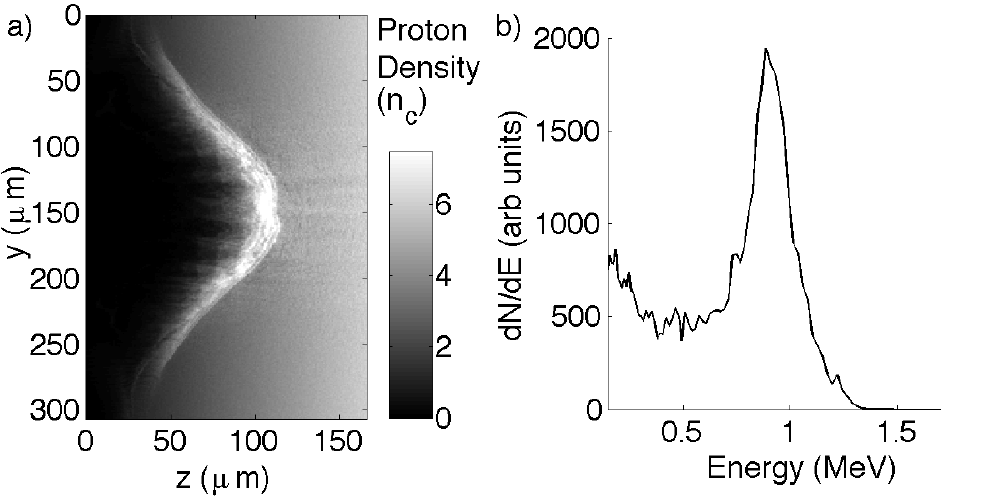}
\caption{{\bf PIC simulation:} (a) Proton density map and (b) proton spectra as seen by simulated spectrometer taken 12ps after the start of the interaction.  Simulation performed  with code  {\sc osiris}, using $20000 \times 3000$ cells to simulate a $600 \times 500\, \mu$m ($\sim 60\times50\, \lambda$) box with laser incident on fully ionised hydrogen plasma with density profile increasing from 0 to 7.5$n_c$ in 100 $\mu$m.  Parameters chosen to replicate the experiment, except with a higher $a_0$ to account for self focusing ($a_0=0.9,  \tau_{L} = 8 \, {\rm ps} = 229$ laser periods, circular polarisation). Both formation of a distinct hole-boring driven shock and an enhanced population of protons at $v=2v_{hb}$ are observed.}
\label{default}
\end{center}
\end{figure}

\pagebreak

\bibliography{bnlRefs}

\begin{methods}

\subsection{Laser parameters}
The experiment used the $\lambda = 10\, \mu$m, 0.5 TW peak power, circularly polarised CO$_{2}$ laser at the Accelerator Test Facility, Brookhaven National Laboratory. The short pulse is achieved in a Kerr cell filled with optically active CS$_{2}$ liquid via fast polarisation switching by a 5 ps long, frequency-doubled ($\lambda = 532$ nm) Nd:YAG beam. Spectral modulation of a picosecond $\lambda = 10 \, \mu$m pulse in CO$_{2}$ gas amplifiers results in splitting of the output pulse into a train of 5~ps pulses with 25 ps period with most of the energy contained in two leading pulses. 
Shots were taken with integrated energy in the range 2.5 - 2.9 J and focused by an $f$/3 off-axis parabolic mirror to a spot size of $w_{0} \approx 70\, \mu$. This gives vacuum target intensities of $6.5-7.5\times 10^{15}\, {\rm Wcm^{-2}}$, or normalised vector-potential  $a_{0}\simeq 0.51 - 0.55$,  where for circular polarisation $\displaystyle a_{0} = {eE \over m\omega c} \simeq 0.60\, (I\lambda^{2})^{1/2}$ (for $I$ in $10^{18}\, {\rm Wcm}^{-2}$ and $\lambda$ in $\mu$m).

\subsection{Experimental set-up}
The laser was focused onto the front surface of a hydrogen gas jet from a $L = 1$~mm circular nozzle. 
A $\tau = 10$ ps,  2$\omega$ ($\lambda = 532$ nm) Nd:YAG laser beam, absolutely synchronised with the CO$_{2}$ beam, was used for probing. The relative timing between driver and probe pulse was varied using an optical delay line. The probe, after passing transversely through the plasma, was split and directed into shadowgraphy and interferometry channels.  These gave information not only about plasma creation and evolution, but also provided the in-situ neutral density profile.  Along the optical axis, $\sim 0.7$ mm above the nozzle edge, the density had an approximately triangular density profile, going from zero to maximum density over a length of $\approx 825\, \mu$m.  The electron (and thus proton) density could be varied up to a maximum of $n_{e} \sim 10^{20}  \, {\rm cm}^{-3} \simeq 10\, n_{cr}$.
The ion beam was characterised with a magnetic spectrometer, which dispersed the protons by their deflection in a transverse magnetic field. The aperture of the spectrometer was a $\phi=0.6$~mm diameter pinhole. The dispersed protons were detected with a (polyvinyltoluene) scintillator screen which light emission was calibrated to the dose of energy deposited by protons impinging on it. The scintillator was reimaged on to a Andor EMCCD camera. 

\subsection{Data unfolding}
Proton energy spectra were unfolded by first removing hot-spots (due to x-rays), background subtracting, and then integrating vertically to give a line-out. An absolute residuals optimisation was used to find the trial spectra which best reproduced the measured signal after convolving with the instrument function, which was taken from the original vertical spread of the signal. We note that due to the small acceptance angle of the spectrometer ($9.8\times10^{-6}$~sr), the transport through the spectrometer has little effect on the signal spread. Density information was obtained from the interferograms, by first obtaining the phase with reference to a background image. An Abel transform was then used to obtain the density profile from the phase change, assuming cylindrical symmetry around the laser axis.

\end{methods}

\begin{addendum}
 \item The work was partly funded by the Libra Basic Technology Consortium and US DOE grant DE-FG02-07ER41488.  We thank; D. Neely, P. Foster and J. Green for providing spectral response calibration of the scintillator, the {\sc Osiris} consortium (UCLA/IST/USC) for use of {\sc Osiris}, K. Kusche and the ATF technical staff for their assistance with the experiment and A. E. Dangor for useful discussions.
 
 \item[Correspondence] Correspondence and requests for materials
should be addressed to Z. Najmudin~(email: zn1@ic.ac.uk).
\end{addendum}

\end{document}